\documentclass[aps,pre,reprint]{revtex4-1}
\usepackage{amsmath,amsfonts,amssymb}
\usepackage{blindtext}
\usepackage{graphicx,float}
\usepackage{epstopdf}
\usepackage[utf8]{inputenc}
\usepackage[encapsulated]{CJK}

\usepackage{mathtools}
\DeclarePairedDelimiter\floor{\lfloor}{\rfloor}

\newcommand{\kpr}{k^{\prime}}
\newcommand{\kavg}{\langle k\rangle}

\newcommand{\changed}[1]{{#1}}
\usepackage[normalem]{ulem} 
\newcommand{\deleted}[1]{}

\begin{document}

\title{Degree Correlations Amplify the Growth of Cascades in Networks} 
\author{Xin-Zeng Wu (\begin{CJK}{UTF8}{bkai}吳信增\end{CJK}) $^{1,2}$} 
\author{Peter G. Fennell$^1$}
\author{Allon G. Percus$^{3,1}$}
\author{Kristina Lerman$^1$}
\affiliation{$^{1}$ Information Sciences Institute, University of Southern California, Marina del Rey, CA 90292}
\affiliation{$^{2}$ Department of Physics and Astronomy, University of Southern California, Los Angeles, CA 90089}
\affiliation{$^{3}$ Institute of Mathematical Sciences, Claremont Graduate University, Claremont, CA 91711}

\begin{abstract}
Networks facilitate the spread of cascades, allowing a local perturbation to percolate via interactions between nodes and their neighbors. We investigate how network structure affects the dynamics of a spreading cascade. By accounting for the joint degree distribution of a network within a generating function framework, we can quantify how degree correlations affect both the onset of global cascades and the propensity of nodes of specific degree class to trigger large cascades. However, not all degree correlations are equally important in a spreading process. We introduce a new measure of degree assortativity that accounts for correlations among nodes relevant to a spreading cascade. We show that 
the critical point defining the onset of global cascades has a monotone relationship to this new assortativity measure. 
In addition, we show that the choice of nodes to seed the largest cascades is strongly affected by degree correlations. Contrary to traditional wisdom, when degree assortativity is positive, 
low degree nodes are  more likely to generate largest cascades. Our work suggests that it may be possible to tailor spreading processes by manipulating the higher-order structure of networks.
\end{abstract}

\maketitle

\section{Introduction}
A local perturbation in a network can spread to many nodes through the interactions of nodes with their neighbors. Such effects have been studied in numerous contexts, including critical phenomena~\cite{Goltsev2008} and percolation~\cite{Karrer2014}. One of the simplest models used to describe these interactions is
the threshold model~\cite{Granovetter78,Watts2002}, where a node changes its state to one held by a sufficiently large fraction of its neighbors. Under some conditions, even a single node can trigger a global cascade that affects a significant portion of the network~\cite{Watts2002}. Despite its simplicity, the threshold model has been used to study a surprisingly wide range of social, biological, and technological phenomena. For example, in social systems, the local perturbation could represent adoption of an innovation, e.g., a new product, action, or idea, that spreads throughout society as individuals adopt the behavior of their friends~\cite{Granovetter78,Watts2002,Centola2010}.
In technological systems, the perturbation could represent the failure of a single component, which triggers a cascade of failures in connected components.
If small shocks can become pandemic, then nodes that can seed global outbreaks will have outsize importance~\cite{Kempe2003,Goltsev2008,Karrer2014}. Such seeds represent influential individuals in social systems, who can help an innovation become widely adopted, or ``weak links'' in technological systems, whose failure compromises the robustness of the entire system. For example, cascading failures were implicated in widespread blackouts in the power grid~\cite{Watts1998}, and market crashes in financial systems~\cite{haldane2011systemic}.

Under what conditions do cascades spread globally? 
Researchers have examined how the dynamics of cascades are affected by network structure, including degree distribution~\cite{Watts2002} and degree correlations between connected nodes~\cite{Gleeson2008,Payne2009}. The latter property is important, because nodes in real-world networks are not connected at random, but tend to link to other nodes with either a similar or dissimilar degree. Following Newman~\cite{Newman2002} network scientists have used {degree assortativity} to measure the correlation of degrees of connected nodes: in positively assortative networks, nodes with similar degrees are connected, e.g., high-degree nodes connected to other high-degree nodes, while in negatively assortative networks, high-degree nodes tend to be connected to low-degree nodes. 
Surprisingly, networks with both sufficiently positive and sufficiently negative assortativity have been found to be vulnerable to global outbreaks.
Such non-monotone behavior was
reported in Erd\H{o}s-R\'enyi random networks~\cite{Payne2009},
and for $k$-core networks~\cite{Gleeson2008}.
However, these works
did not explain the anomalous relationship between degree assortativity and cascade size, nor did they provide a general mathematical framework for quantifying how degree correlations affect the properties of cascades under the threshold model.
Our work addresses these topics and also demonstrates that assortativity does not adequately capture some important aspects of degree correlations in networks.
Specifically, in real-world networks, assortativity is heavily skewed by nodes with a single neighbor, which act as ``dead ends'' to spreading cascades. To address this shortcoming, we introduce a new measure, which accounts for degree correlations of nodes that participate in a spreading cascade. We demonstrate that cascade size is monotonically related to this measure.



In this paper, we study dynamics of cascades that spread on networks according to the Watts threshold model~\cite{Watts2002}. Every node in this model can be in one of two possible states: either active or inactive (e.g., adopting an innovation or not). Nodes change their state based on the states of their neighbors, with activated nodes remaining active in the later updating steps.
Specifically, an inactive node with $k$ neighbors  becomes active if at least $\phi k$ of its neighbors are active.
The activation threshold $\phi$ takes values $0\leq\phi\leq1$, with smaller values of $\phi$ rendering a node more susceptible to the influence of neighbors. 
Following Watts, we call a node \emph{vulnerable} if it can be activated by a single active neighbor. In a threshold model
this is equivalent to a node of degree $k$ having an activation threshold $\phi < 1/k$.
We study undirected networks with degree distribution $p_k$, giving the probability that a randomly chosen node has degree $k$, and joint degree distribution $e_{k,\kpr}$, giving the probability that a randomly chosen edge in the network links nodes of degrees $k$ and $\kpr$~\cite{Newman2002}.
We ignore correlations beyond two neighboring nodes, so structures of higher-order than $e_{k,\kpr}$ are assumed to be random.
The joint degree distribution matrix $e_{k,\kpr}$ is symmetric and related to $p_k$ through $\sum_{k'=1}^{\infty}e_{k,\kpr} = q_k = kp_k/\langle k \rangle$, where $q_k$ here is the probability that a node at the end of a randomly chosen link has degree $k$, and $\langle k \rangle$ is the network's average degree.
Globally, the strength of degree correlation in an undirected network can be quantified by the assortativity coefficient: 
\begin{equation}
\label{eq:assortativity}
r=\frac{1}{\sigma_{q}^2}\sum_{k,\kpr}k\kpr \left[e_{k,\kpr}-q_k q_{\kpr}\right].
\end{equation}
Here, $\sigma^2_q= \sum_k{k^2 q_k} - \left[ \sum_k{k q_k}\right]^2$. In assortative (resp. disassortative) networks with $r > 0$ (resp. $r<0$), nodes have a tendency to link to similar (resp. dissimilar) degree nodes, e.g., high-degree nodes to other high-degree (resp. low-degree) nodes.

We use the generating function approach~\cite{Dodds2009} to derive the expected size of cascades triggered by a single active node. In the subcritical regime, cascades never reach an appreciable fraction of the network, but when the system transitions to the supercritical regime, global cascades are possible. We derive the condition for this supercritical transition to occur. The supercritical formulas yield the expected size of cascades given the seed's degree. To better understand how degree correlations  affect cascades, we model the joint degree distribution using a bivariate log-normal distribution~\cite{Wu2017}.  Strong assortative behavior renders networks vulnerable to global outbreaks. Surprisingly, the same holds for strongly disassortative behavior, as long as there are enough links between vulnerable nodes. This highlights the limits of using assortativity to measure degree correlation.  Additionally, we show that in some networks when assortativity is strongly positive (or negative), lower degree nodes are more influential, as they can trigger larger outbreaks. As assortativity approaches zero from either direction, influence shifts to higher degree nodes. This phenomenon can be explained by the fraction of edges that link vulnerable nodes.

We replicate these findings by simulating cascades on synthetic random networks with power-law degree distribution, 
which have been rewired to obtain a range of degree-degree correlations. Both the onset of global outbreaks and their size agree with theory. However, in real-world networks drawn from diverse domains, cascade size is systematically smaller than theoretical predictions. This reflects the fact that real-world networks have structure beyond that given by degree-degree correlations~\cite{Gleeson2008,Wu2017}. After rewiring the networks so as to eliminate the higher-order structure, while preserving the degree and joint degree distributions, we find the agreement with theoretical predictions of simulated outbreak sizes restored. This suggests that higher-order network structure suppresses outbreaks. Despite this, theory predicts well who the influential nodes are.

\section{Theory}
\subsection{Subcritical cascades}
Consider a network with $N$ nodes, in the limit where $N\to\infty$.
We adopt the setting of~\cite{Dodds2009}, distinguishing between a {\em
local\/} cascade that reaches at most a vanishing fraction of nodes
(such as some countable number), and a {\em
global\/} cascade that spreads to a nonvanishing fraction of the nodes.
Given a seed node of degree $k$, we denote by $\pi_{k,n}$ the probability
that it generates a local cascade, of size $n$, and we denote by
$g_k$ the probability that it generates a global cascade.
We represent the distribution 
$\pi_{k,n}$ through its generating function $H_{0,k}(x) = \sum_{n=0}^\infty\pi_{k,n}x^n$.
Generating functions easily allow  calculating many key properties of
cascades, such as the total probability of a local cascade $H_{0,k}(1) = \sum_{n=0}^\infty \pi_{k,n}$ or its mean size  $H_{0,k}'(1) = \sum_ {n=0}^\infty n \pi_{k,n}$.
Since a seed node can generate either a local or a global cascade,
$H_{0,k}(1) + g_k = 1$.

First, we consider the \emph{subcritical regime}, where only local cascades exist and $g_k=0$. In this regime, the size of a local cascade generated by a seed node of degree $k$ can be decomposed as one (the seed itself) plus the collective size of cascades generated by its $k$ neighbors. We denote as $H_{1,k}(x)$ the generating function for the size of the local cascade created by the neighbor of a $k$-degree node.  The power-rule of generating functions~\cite{Newman2002} states that if $k$ independent realizations of a random process with generating function $G(x)$ are created, then the probability distribution for the sum of the outcomes has generating function $[G(x)]^k$.  Therefore,
\begin{equation}
	H_{0,k}(x) = x[H_{1,k}(x)]^k,
	\label{eq:H0}
\end{equation}
where the leading factor $x$ on the right hand side of Eq.~\eqref{eq:H0} has the effect of adding one to the combined size of the cascade generated by the neighbors of the seed node (and thus accounting for the seed node itself). To determine $H_{1,k}(x)$, we first note that the neighbor of the $k$-degree seed node can only generate a cascade if it itself is vulnerable, i.e., it can be activated by a single active neighbor. This will occur if the neighbor's degree $k'$ satisfies $\kpr \leq \phi^{-1}$. Given that the neighbor is activated, the size of the cascade generated by that neighbor is one plus the size the cascade generated by its neighbors down the tree (that is, not including the seed). Hence, $H_{1,k}(x)$ is given by
\begin{multline}
  H_{1,k}(x) = x\frac{e_{k,1}}{q_k} + \sum_{k'>\phi^{-1}}\frac{e_{k,\kpr}}{q_k} \\+ x \sum_{k'=2}^{\floor*{\phi^{-1}}}\frac{e_{k,\kpr}}{q_k} [H_{1,k'}(x)]^{k'-1},
  \label{eq:H1}
\end{multline}
where the first term on the right hand side of Eq.~\eqref{eq:H1} is the probability that the neighbor of a degree-$k$ node has no other outgoing links (i.e., has degree $\kpr=1$), the second term is the probability that the neighbor is not vulnerable, and the third term is the generating function for the size of the cascade generated from its $\kpr-1$ down-tree neighbors. The right hand side of Eq.~\eqref{eq:H1} only contains $H_{1,\kpr}(x)$ terms, and indeed if we were to consider the next level of the tree (i.e., neighbors of the seed's neighbors) then we would obtain the same equations, as our network model only considers correlations between nearest neighbors. Thus, the full
system of equations describing the distribution of the sizes of local cascades generated by a degree-$k$ seed is given by Eqs.~\eqref{eq:H0} and \eqref{eq:H1} for $k_{\min} \leq k \leq k_{\max}$.

\subsection{The Onset of Global Cascades}

In the subcritical regime, the mean cascade size $\langle s_{k_0} \rangle$ generated by a $k_0$-degree seed node is $H'_{0,k_0}(1)$ (as $g_k=0$). Differentiating Eqs.~\eqref{eq:H0} and \eqref{eq:H1} and evaluating at $x=1$ gives a system of equations for $\langle s_{k_0} \rangle$,
\begin{align}
  \langle s_{k_0} \rangle &= 1 + k_0H'_{1,k_0}(1), \label{eq:mean_cascade_size1} \\
  H_{1,k}'(1) &= \sum_{k'=2}^{\floor*{\phi^{-1}}}\frac{e_{k,\kpr}}{q_k}\left[1 + (k'-1)H'_{1,k'}(1)\right],
  \label{eq:mean_cascade_size2}
\end{align}
where we have used the identity $H_{0,k}(1)=H_{1,k}(1)=1$ in the subcritical regime. Eq.~\eqref{eq:mean_cascade_size2} holds for all values of $k$, and can be written in matrix form as $\mathbf{h}'_1(1) = \mathbf{B}\mathbf{1} + \mathbf{B}\mathbf{U}\mathbf{h}'_1(1)$.  Here, $\mathbf{h}'_1(1)$ is the state vector with $(\mathbf{h}'_1(1))_k = H_{1,k+1}'(1)$, $\mathbf{B}$ is a matrix with elements $\mathbf{B}_{kk'}=e_{k,\kpr}/q_k$ for $2\leq k' \leq \floor*{\phi^{-1}}$ and 0 otherwise, $\mathbf{1}$ is a vector of ones, and $\mathbf{U}$ is a diagonal matrix with elements $\mathbf{U}_{kk} = (k-1)$.
\changed{Notice from the upper limit of the sum in Eq.~\eqref{eq:mean_cascade_size2} that the threshold $\phi$ affects the system of equations discretely, through the integer value $\floor*{\phi^{-1}}$.}

The linear system can be rearranged to give
\begin{equation}
  (\mathbf{I} - \mathbf{B}\mathbf{U})\mathbf{h}'_1(1) = \mathbf{B}\mathbf{1}.
  \label{eq:mat_eq}
\end{equation}
In the subcritical regime, $\mathbf{I} - \mathbf{B}\mathbf{U}$ can be inverted to give an expression for $\mathbf{h}'_1(1)$ that is positive and finite. However, there exists a critical \changed{value of $\phi$} marking the onset of global cascades where  $\mathbf{I} - \mathbf{B}\mathbf{U}$ is no longer invertible, and hence there is no solution to Eq.~\eqref{eq:mat_eq}. This critical point can be expressed formally as
\begin{equation}
  \det(\mathbf{I} - \mathbf{B}\mathbf{U}) = 0.
  \label{eq:threshold_condition}
\end{equation}
The critical point marks 
the transition to the \emph{supercritical regime}, where the network is vulnerable to global outbreaks (i.e., $g_k \ne 0 $). Note that Eq.~\eqref{eq:threshold_condition} corresponds to the largest eigenvalue of $\mathbf{B}\mathbf{U}$ being equal to one, and from Eq.~\eqref{eq:mat_eq} we can see that the supercritical regime exists if and only if the largest eigenvalue of $\mathbf{B}\mathbf{U}$ is greater than one.  \changed{Since the critical threshold is the largest value of $\phi$ for which this condition holds, it
will be the inverse of an integer degree.} 

In general, the critical condition depends on the threshold $\phi$, the degree distribution $p_k$, and the joint degree distribution $e_{k,\kpr}$ through the matrix $\mathbf{B}$. Thus, for a given degree distribution and threshold, degree-degree correlations determine whether the network is vulnerable to global outbreaks. To illustrate this, we consider two idealized cases: a network in which the degrees of nodes are completely independent, and one in which they are perfectly correlated. The first case is a network with no degree-degree correlations; therefore, $e_{k,\kpr} = q_k q_{\kpr}$.
\changed{As a consequence of Eq.~\eqref{eq:threshold_condition},}
the critical point exists when
\begin{equation}
	\sum_{k=2}^{\floor*{\phi^{-1}}}\frac{k p_k}{\langle k\rangle}(k-1) = 1.
	\label{eq:H1_uncorr}
\end{equation}
The sum on the left hand side of Eq.~\eqref{eq:H1_uncorr} is a decreasing function of the threshold $\phi$, and if the degree distribution satisfies $\sum_{k=1}^{\infty}k^2 p_k/\langle k \rangle > 2$, then there will always exist a critical threshold $\phi^*$ for which global cascades can occur for values $\phi\leq\phi^*$ but will never occur for $\phi>\phi^*$. Moreover, in this scenario $H_{1,k}'(1)$ is independent of $k$ (Eq.~\eqref{eq:mean_cascade_size2}), and thus from Eq.~\eqref{eq:mean_cascade_size1} we can see that nodes of any degree can generate global cascades in the supercritical regime.

The second case we consider is a perfectly assortative network in which nodes are connected only to other nodes of the same degree: $e_{k,\kpr} = \delta(k,\kpr)q_{\kpr}$, where $\delta(k,\kpr)=1$ if $k=\kpr$ and 0 otherwise.  \changed{While this limiting case may seem artificial in that it results in a collection of disconnected regular graph components (note that more than one degree value is needed in order for assortativity to be well-defined), it allows} Eq.~\eqref{eq:mean_cascade_size2} to reduce to the set of independent equations
\begin{equation}
	\;\; H_{1,k}'(1) =
	\begin{cases}
		1 + (k-1)H_{1,k}'(1) &  2 \leq k \leq \floor*{\phi^{-1}} \\
		0 &  k=1 \mbox{ or }k > \floor*{\phi^{-1}}
	\end{cases}
	\label{eq:H1_assort}
\end{equation}
which, rearranged, gives $H_{1,k}'(1) = 1/(2-k)$ for $2 \leq k \leq \floor*{\phi^{-1}}$. This expression is either infinite or negative for all values of $k$, and since $H_{1,k}'(1)$ should be positive and $H_{1,k}'(1)=0$ is not a solution to Eq.~\eqref{eq:H1_assort}, the only possible solution for these degree classes $k$ is the global cascade condition $H_{1,k}(1)\rightarrow \infty$. Thus, in perfectly assortative networks, there will always be global cascades when $\phi \leq 1/2$, but from Eq.~\eqref{eq:mean_cascade_size1} we see that these cascades can only be triggered by seed nodes that are vulnerable (i.e., $2 \leq k \leq \floor*{\phi^{-1}}$).

The two scenarios above illustrate not only that degree-degree correlations affect the onset of global cascades in networks, 
but also that such correlations affect the ability of seeds in different degree classes to trigger global outbreaks.



\subsection{Supercritical Regime and  Node Influence}

To identify influential nodes, we calculate the expected size of the cascades they trigger. A node can be considered influential if it is able to seed a global outbreak~\cite{Kempe2003}. In the supercritical regime, the network is composed of a single vulnerable component
, and thus a node will trigger a global outbreak if and only if it belongs to this giant vulnerable component (GVC). The size $S$ of the giant vulnerable component can be calculated as follows. A $k$-degree node will be in the GVC if and only if it triggers a global cascade, and so the fraction of $k$-degree nodes that are in the GVC is the fraction that will generate global cascades, i.e., $g_k$. Summing over all degree classes, the size of the giant vulnerable component is
\begin{equation}
	S = \sum_{k=1}^{\infty}p_k g_k.
\end{equation}
Because a $k$-degree node either belongs to the GVC or not, the expected fraction of the network triggered by a $k$-degree seed is $S_k=S$ for nodes in the GVC, and $S_k=0$ for nodes outside of the GVC, and thus we have that $S_k = Sg_k$. Then, as $g_k = 1-H_{0,k}(1)$, and $H_{0,k}(1) = [H_{1,k}(1)]^k$ (from Eq.~\eqref{eq:H0}), we have a system of equations that give us the expected outbreak size triggered by a node of degree $k$:
\begin{align}
  S_k =& \left(1 - \left[H_{1,k}(1)\right]^k\right)S \label{eq:S_k0}\\
  H_{1,k}(1) =& \frac{e_{k,1}}{q_k} + \sum_{k'>\phi^{-1}}\frac{e_{k,\kpr}}{q_k} \nonumber \\
  &\hspace{0.2in} + \sum_{k'=2}^{\floor*{\phi^{-1}}}\frac{e_{k,\kpr}}{q_k} [H_{1,k'}(1)]^{k'-1} \label{eq:H_1_x=1} \\
  S =& \sum_{k'=1}^{\infty}p_{k'}\left(1 - \left[H_{1,k'}(1)\right]^{k'}\right) \label{eq:S}
\end{align}
Equation~(\ref{eq:H_1_x=1}), for all values of $k$, gives a set of polynomial equations that can be solved for $H_{1,k}(1)$, and substituting these values into Eqs.~\eqref{eq:S_k0} and \eqref{eq:S} gives us the expected cascade size $S_k$ triggered by nodes of degree $k$.


\section{Bivariate Log-normal Model}
To understand better how the properties of cascades depend on network structure, we model $e_{k,\kpr}$ as a bivariate log-normal distribution~\cite{Wu2017} with parameters ($\mu,\sigma^2,c$), where $\mu$ is the location parameter, $\sigma^2$ is the scale parameter, and $c$ is the correlation coefficient.
\begin{equation}
\log \kpr|\log k\sim\mathcal{N}\left[\mu+c(\log k-\mu),(1-c^2)\sigma^2\right].
\end{equation}
The assortativity of the network is then
\begin{equation}
r = \frac{\mathrm{Cov}(k,\kpr)}{\mathrm{Var}(k)}=\frac{e^{c\sigma^2}-1}{e^{\sigma^2}-1}.
\end{equation}
Note that, for a given value of $\sigma^2$, the assortativity is bounded by $-e^{-\sigma^2}\leq r\leq 1$.

\begin{figure}[h]
    \centering
    \includegraphics[width=0.47\textwidth]{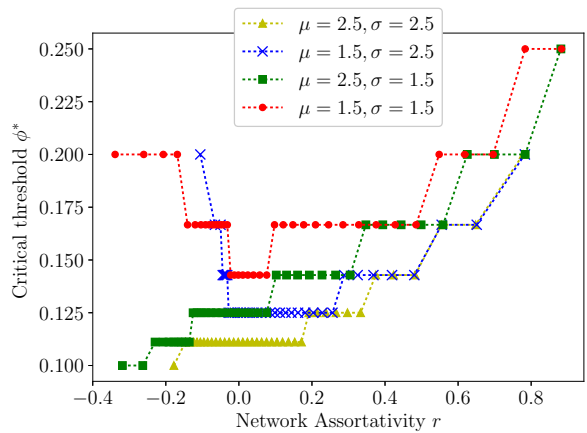}
    \caption{(Color online) 
   Critical threshold $\phi^*$ as a function of network's degree assortativity. Global cascades exist in the parameter region $\phi < \phi^*$, and only local cascades exist in the region $\phi > \phi^*$.
    }
    \label{onset}
\end{figure}

We identify the location of the critical point defining the onset of global cascades by inserting the log-normal distribution into Eq.~\eqref{eq:threshold_condition} and varying $c$ to produce a range of assortativity values.
Fig.~\ref{onset} shows the series of critical points obtained from the theory using parameters $\mu=1.5$ and 2.5, $\sigma=1.5$ and 2.5 as a function of assortativity $r$. The critical point is defined by the threshold $\phi^*$, such that
global cascades exist in the parameter region $\phi < \phi^*$, and only local cascades exist in the region $\phi > \phi^*$. A maximum degree in the joint degree distribution needs to be introduced here in order to solve the outbreak size by Eq.~\eqref{eq:H_1_x=1}. We set the cutoff to be $k_\mathrm{max}=1,000$. As expected, assortative ($r>0$) networks are more vulnerable to global outbreaks, 
since with increasing assortativity, vulnerable nodes are more likely to connect to other vulnerable nodes, which helps create the giant vulnerable component on which global outbreaks spread.

Interestingly, as observed by Payne et al.~\cite{Payne2009}, disassortative ($r<0$) networks are vulnerable to global outbreaks as well.  Fig.~\ref{onset} shows that, for certain values of $\mu$ and $\sigma$, the critical point $\phi^*$ displays non-monotone behavior and in fact increases with disassortativity.  At first this may appear puzzling, since in the disassortative regime, vulnerable nodes are less likely to connect to other vulnerable nodes, which would seem to inhibit the formation of a giant vulnerable component on which cascades spread.  That intuition is flawed, however, because of
the heterogeneous nature of the joint degree distribution:
even in networks that are globally disassortative, nodes of lower degree can be linked assortatively.  Some numerical evidence of this has been seen in~\cite{Payne2009} for the $r=-0.8$ case of a random network, where nodes with degrees $k=2$ through $k=4$ have a strong tendency to link to one another.


\begin{figure}[h]
\centering
\includegraphics[width=0.47\textwidth]{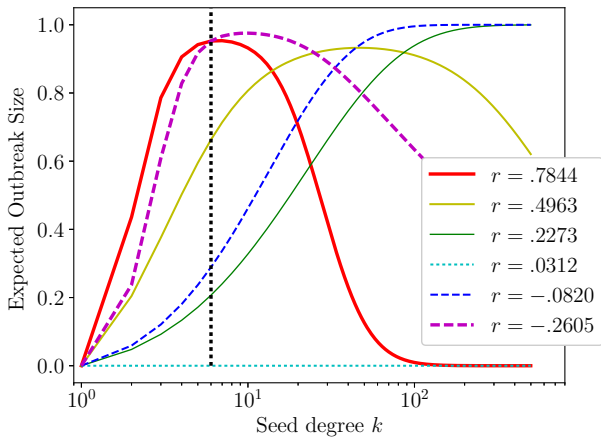}
\caption{(Color online) 
Expected outbreak size $S(k)$ on network with log-normal distribution network joint degree distribution $e_{k,\kpr}$. The distribution is with parameters $\mu=1.5$ and $\sigma=1.5$, and the assortativity parameter $c$ is tunable between range $-1$ and 1. The vertical dashed line indicates the threshold chosen $\phi=\frac{1}{6}$.}
\label{lognormal}
\end{figure}

We also consider the impact of assortativity on how influential nodes of a given degree are, as measured by their ability to initiate large cascades. 
Fig.~\ref{lognormal} shows the expected size of cascades triggered by seeds in different degree classes. 
Perhaps surprisingly, the best-connected nodes---hubs---are not always the most influential. Instead, we again see a non-monotone effect. For highly assortative networks, the largest cascades are triggered by lower-degree nodes. As assortativity decreases, influence shifts to higher-degree nodes, and lower-degree nodes lose their ability to trigger large outbreaks. But when assortativity becomes negative, the picture changes: influence shifts back to lower degree nodes. In highly disassortative networks, similar to the highly assortative networks, the degree of the most influential nodes is not far from the inverse of the threshold ($\phi^{-1}$).


\begin{figure}[h]
    \centering
    \includegraphics[width=0.47\textwidth]{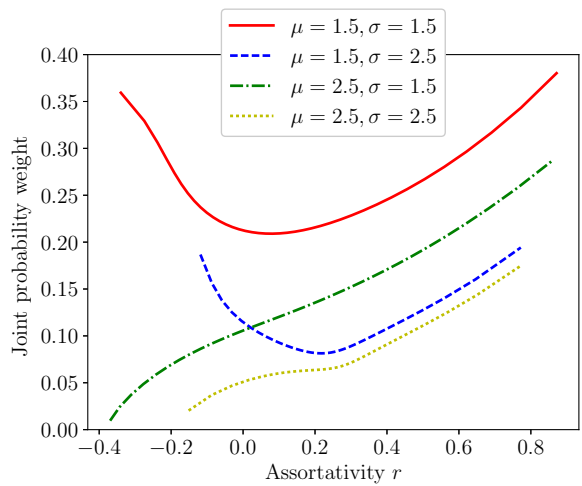}
    \caption{(Color online) The block probability weight on different threshold chosen in $e_{k,\kpr}$ with $\mu=1.5,2.5$ and $\sigma=1.5,2.5$ which sums over the element values in range of $2\leq k\leq\phi^{-1}_0$, $2\leq \kpr\leq\phi^{-1}_0$, where $\phi^{-1}_0$ solves the equation of Eq.~\eqref{eq:H1_uncorr}.}
    \label{block}
\end{figure}

In order to explain these effects, consider how 
degree correlations among nodes in disassortative networks can promote global outbreaks.
From Eq.~\eqref{eq:threshold_condition}, it is clear that the matrix $\mathbf{BU}$  dictates the occurrence of global outbreaks. This matrix has entries $e_{k,\kpr} (\kpr-1)/q_k$ for $2\leq \kpr \leq \floor*{\phi^{-1}}$, and as $\phi$ decreases, the nonzero part of the matrix increases. At some point, it may reach a size for which the largest eigenvalue is greater than one, at which point it will be unstable to global outbreaks. The interpretation of this is that decreasing $\phi$ makes an increasing number of nodes vulnerable, and when a sufficient number of links exist between vulnerable nodes --- or when a sufficient amount of probability mass is contained in the submatrix $e_{k,\kpr}$ for $2\leq k, \kpr \leq \floor*{\phi^{-1}}$ --- then a giant vulnerable component will form.
Fig.~\ref{block} shows how the probability mass of $e_{k,\kpr}$ in the vulnerable regime (i.e., $e_{k,\kpr}$ for $2\leq k, \kpr \leq \floor*{\phi^{-1}}$) changes as a function of the global assortativity coefficient. 
Comparing this plot to Fig.~\ref{onset}, we can see that the parameter regime where the block probability mass is large corresponds to networks that are more unstable to global outbreaks, even when they are disassortative at the global level.

\begin{figure}[h]
    \centering
    \includegraphics[width=0.47\textwidth]{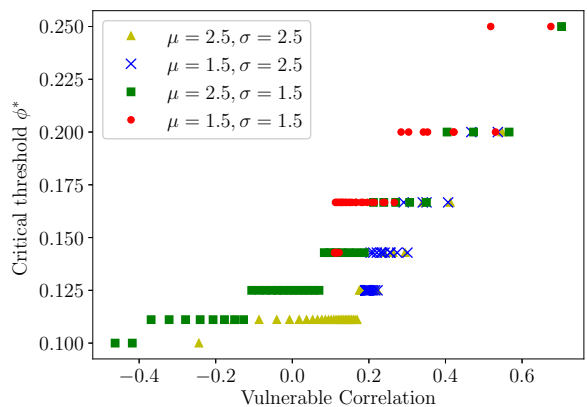}
    \caption{(Color online) 
   Critical threshold $\phi^*$ as a function of the network's vulnerable node degree correlation defined by Eq.~\eqref{eq:vul_corr}. Global cascades exist in the parameter region $\phi < \phi^*$, and only local cascades exist in the region $\phi > \phi^*$.
    }
    \label{onset2}
\end{figure}


As an alternative to assortativity, we propose a measure that quantifies degree correlation between vulnerable nodes. We introduce an indicator for node $i$ to be vulnerable as $v_i=\boldsymbol{1}_{2\leq k_i\leq\floor*{\phi_0^{-1}}}$, where $\phi_0^{-1}$ solves Eq.~\eqref{eq:H1_uncorr}. Here $P(v_i=1)=\sum_{k=2}^{\floor*{\phi_0^{-1}}}q_k=Q_v$. Then the correlation between degrees of vulnerable nodes is

\begin{equation} \rho_{\mathrm{vul}}=\frac{\sum_{k=2}^{\floor*{\phi_0^{-1}}}\sum_{\kpr=2}^{\floor*{\phi_0^{-1}}}e_{k,\kpr}-Q_v^2}{Q_v(1-Q_v)}.
	\label{eq:vul_corr}
\end{equation}
When we plot the critical threshold as a function of this new measure---vulnerable node degree correlation---we observe a monotone behavior (Fig.~\ref{onset2}). That suggests that this quantity, rather than degree assortativity, is an appropriate control parameter determining the size of outbreaks.  Both the most disassortative and the most assortative networks in Fig.~\ref{onset} have high vulnerable correlation in Fig.~\ref{onset2}.  This also explains the effect seen in Fig.~\ref{lognormal}.  While high-degree nodes can influence more nodes once they are activated, low-degree nodes are more likely to be vulnerable.  Thus, when the vulnerable node degree correlation is high, large cascades are more likely to be initiated by nodes of lower degree.

\section{Results}
We use the theoretical framework to study vulnerability of synthetic and real-world networks to global outbreaks and the properties of cascades in such networks.

\subsection{Cascades in Synthetic Networks}
\changed{Using the configuration model, we generate networks with $N=10,000$ nodes and a power law-like degree sequence with exponents $\alpha=2.1$ and 2.4. The resulting networks have assortativity $r=-0.17$ and $-0.07$ respectively.  We then change the degree correlations in the networks by rewiring them according to Newman's method~\cite{Newman2002}. The rewiring procedure picks two pairs of linked nodes at random and exchanges their edges if doing so changes the assortativity in the desired direction. Note that this procedure only changes the so-called \emph{2K structure} of the network~\cite{Wu2017}---its joint degree distribution $e_{k, \kpr}$ matrix---without changing its degree distribution. Through this process, we obtain a series of networks that share the same degree sequence but span a range of assortativity values from  $r=-0.1$ to $-0.3$ for $\alpha=2.1$, and $r=0.05$ to $-0.15$ for $\alpha=2.4$. Note that we choose the degree sequence with not too large maximum degree to allow greatest flexibility for the rewiring process.}
%

\begin{figure}[h]
\centering
\includegraphics[width=0.47\textwidth]{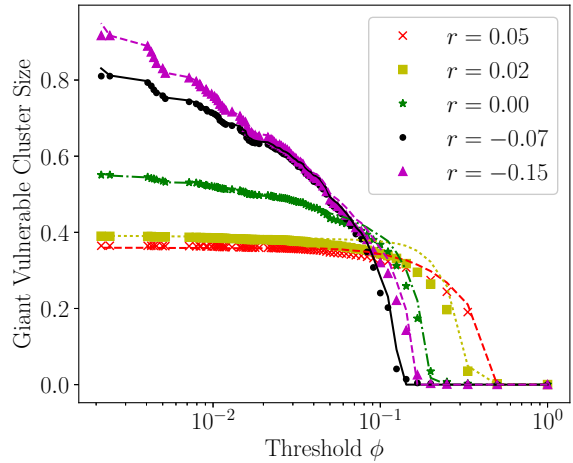}
\caption{(Color online) Theoretical calculation (lines) and observations (points) of giant vulnerable component of synthetic power-law networks with exponent $\alpha=2.4$ and number of nodes $n=10,000$.}
\label{giantvul_syn}
\end{figure}

Fig.~\ref{giantvul_syn} reports the size of the giant vulnerable component for different values of the threshold $\phi$. This was calculated by plugging the $e_{k,\kpr}$ matrix of the network into the nonlinear equations Eq.~\eqref{eq:H_1_x=1}. 
The critical point signaling the onset of global outbreaks occurs at the inflection point, where the line departs from the x-axis. As assortativity of the rewired network increases ($r>-0.07$), the critical point shifts to ever larger values of $\phi$. 
Degree correlations destabilize the network, creating conditions for outbreaks to spread. This is because in assortative networks, similar degree nodes are connected, with vulnerable nodes more likely to be linked to one another, forming a giant vulnerable component on which outbreaks spread. However, as assortativity increases, the network also becomes more fragmented.  Since cascades are limited to the giant component of the network, assortativity decreases their maximum size. In contrast, as the network becomes more disassortative ($r<-0.07$), we also observe that the upper bound of the onset of global outbreaks increases. \changed{This non-monotonicity is similar to that observed in numerical experiments with the log-normal distribution (Fig.~\ref{onset}). Although the assortativity of the unrewired network (black line, $r=-0.07$) is slightly negative due to the structural cutoff~\cite{Boguna2004}, this network corresponds to the neutral assortativity networks in Fig.~\ref{onset} with lowest values of critical threshold. When the network is more assortative or more disassortative, the critical threshold for the onset of global cascades shifts to larger values, just as in Fig.~\ref{onset}.} 


\begin{figure*}
\centering
\includegraphics[width=0.66\textwidth]{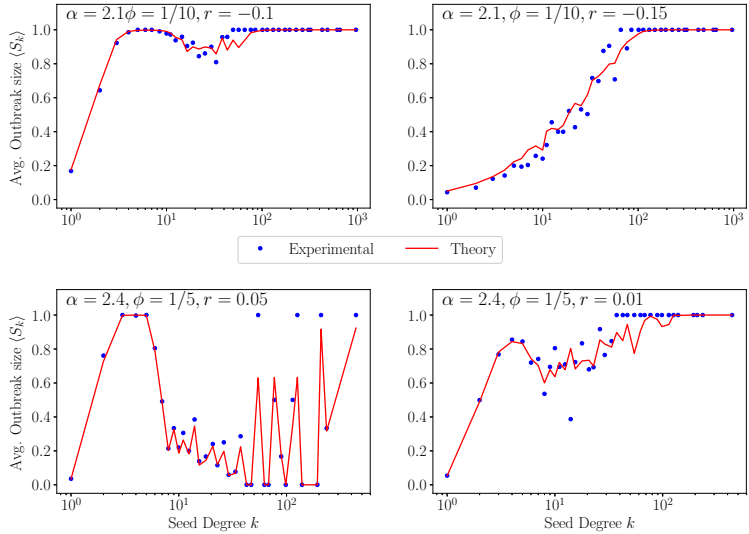}
\caption{(Color online) Theoretical calculation and observations of single seed cascading problems on synthetic power-law networks with $\alpha=2.1$ and 2.4, comparing the observed expected outbreak by the seed degree (blue dots) and the predictions of the generating function formulation (red line).}
\label{powerlaw}
\end{figure*}

Next, we study node influence, which, as explained earlier, we measure by its ability to trigger large cascades. \changed{We simulate cascades using the threshold that puts the network in a slightly supercritical regime.} Fig.~\ref{powerlaw} shows expected size of cascades predicted by theory, together with the average size of the \changed{simulated} outbreaks triggered by nodes of different degree. Each plot in Fig.~\ref{powerlaw} with $\alpha=2.4$  produces a single point in Fig.~\ref{giantvul_syn} via a weighted sum by network degree distribution. Again, the observations made in synthetic networks are similar to those for the log-normal model.

In the more disassortative networks (right hand plots in Fig.~\ref{powerlaw}), the high-degree nodes are more influential. However, when assortativity increases, low-degree nodes also become influential. 
In the more assortative networks (left hand plots in Fig.~\ref{powerlaw}), both the low-degree and high-degree nodes initiate larger cascades, on average, than the middle-degree nodes.

\subsection{Cascades in Real-World Networks}
\begin{figure}[h]
\centering
\includegraphics[width=0.47\textwidth]{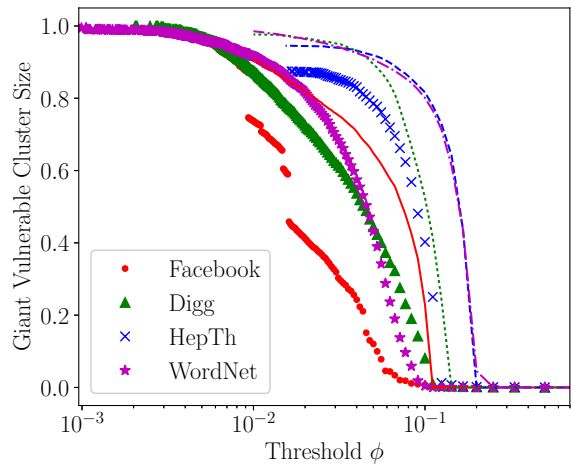}
\caption{(Color online) Theoretical calculation (lines) and observations (points) of giant vulnerable component of selected real-world networks.}
\label{giantvul_real}
\end{figure}
Finally, we study cascade dynamics on real-world networks, which include biological, social, and semantic networks, ranging in size from 4k to 150k nodes. The basic properties of these networks are listed in Table~\ref{network}. Their assortativity ranges from mildly disassortative ($r=-0.0623$) to strongly assortative ($r=0.6322$). Fig.~\ref{giantvul_real} shows the size of the giant vulnerable component as a function of threshold $\phi$. Unlike in synthetic networks, the theory (lines) does not agree well with results of simulations (symbols). 

\begin{figure*}
\centering
\includegraphics[width=\textwidth]{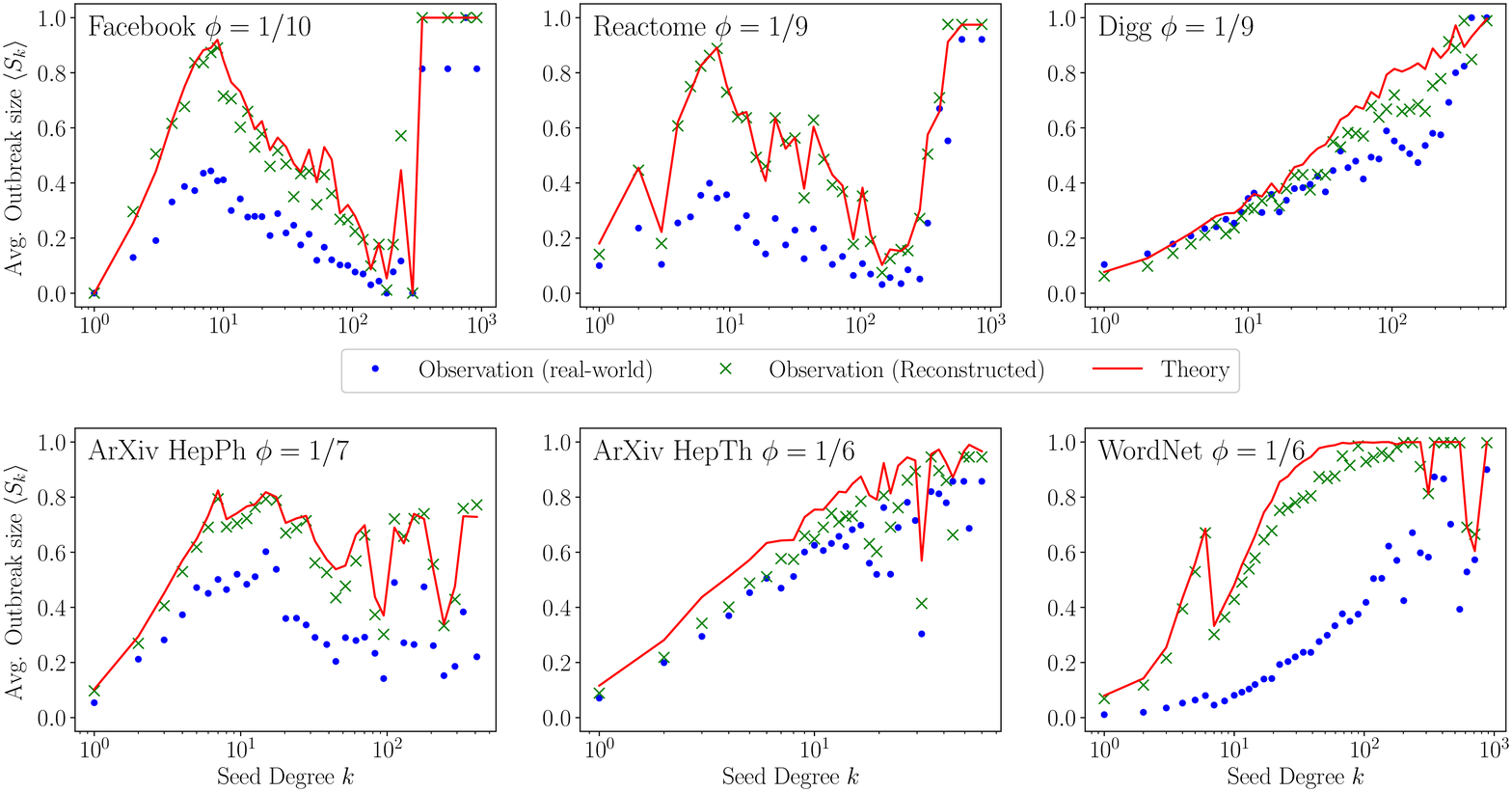}
\caption{(Color online) Theoretical calculation and observations of single seed cascading problems on real-world networks, comparing the observed expected outbreak by the seed degree (blue dots), predictions of the generating function formulation (red line) and the observation on the reconstructed graph (green crosses).}
\label{realworld}
\end{figure*}

We simulate cascading dynamics on real world-networks and measure their size. \changed{The thresholds for these simulations are chosen slightly below the theoretical global cascading threshold by Eq.~\eqref{eq:threshold_condition}.} 
As shown in Fig.~\ref{realworld}, the average size of cascades triggered by seeds of a given degree (blue dots in Fig.~\ref{realworld}) is smaller than theoretical predictions (red lines). This is likely because real-world networks have structure beyond that given by the joint degree distribution $e_{k,\kpr}$, which our theory does not take into account\changed{, and which confines cascades within portions of the network}.
To test the hypothesis, we rewire the network in such a way as to preserve the $e_{k,\kpr}$ and $p_k$ distributions, but destroy higher-order structure \changed{ such as clustering, community structure, or neighbor degree correlations}~\cite{Wu2017}.
In each step of the rewiring, we randomly choose two edges subject to the constraint that one endpoint from each edge shares the same degree value. Then we swap the edges so that each of those endpoints instead links to the other node~\cite{Mahadevan2006}. This step does not change the two edges' contribution toward the joint degree distribution $e_{k,\kpr}$. The step is then implemented a large number of times, roughly equal to the total number of edges. Since the edge chosen depends only on the degrees of its endpoints, the procedure is sufficient to eliminate the higher-order structures in the network.
The average cascade size on such rewired networks (green dots in Fig.~\ref{realworld}) is in agreement with theoretical predictions.
This suggests that higher-order structure beyond degree-degree correlations suppresses outbreaks in real-world networks.

At the same time, we notice that the theory does accurately predict the seed degree at which the outbreak size peaks in these real-world networks.  We explain this as follows.  In the log-normal joint distribution model, we have seen that low-degree nodes (degree near $\phi^{-1}$) are influential when the network is both highly assortative and highly disassortative, and a similar peak phenomenon also exists in assortative synthetic networks.  In real-world networks, on the other hand, the presence of higher-order structure means that there is a mixing of more assortative and more disassortative network elements.  
Our analysis of the log-normal model illustrates that both of these elements, assortative and disassortative, can contribute to a peak of influence around $\phi^{-1}$.
Thus, even if the theory does not reproduce the size of cascades well in real-world networks, it correctly identifies the location of the peak and thereby identifies the influential nodes in the network.

\section{Conclusion}

In this paper, we have explored how the structure of networks affects the dynamics of cascades. We have used a tree-like approximation to calculate the expected size of cascades spreading on networks according to the Watts threshold model. The mathematical formulation allows us to explicitly model the impact of degree correlations, specified by the joint degree distribution $e_{k,\kpr}$, on the 
size of outbreaks triggered by a single node. 
Global outbreaks are more likely in strongly assortative networks, where the degrees of connected nodes are highly correlated. In such networks, nodes that are vulnerable to changing state tend to be connected to other vulnerable nodes, forming a giant connected component on which cascades spread. Surprisingly, strongly disassortative networks are also unstable to global outbreaks, but only when enough vulnerable nodes are connected. Outside of this important block of assortativity, the probability mass of the joint degree distribution matrix is dominated by non-spreading nodes, leading to an overall negative assortativity.  We have introduced a new measure --- vulnerable node degree correlation --- that better captures the size of outbreaks of the Watts threshold model. 

We have also explored the role of seed node degree in cascades.  While low-degree nodes are the most vulnerable, high-degree nodes are typically the most influential as they can trigger the largest outbreaks. On the other hand, in sufficently assortative as well as sufficiently disassortative networks, the low degree nodes turn out to be the most influential.  
We have found that this, too, relates closely to the correlation between degrees of vulnerable nodes.  When that correlation is sufficiently high, which corresponds to both the highly assortative and the highly disassortative case, the vulnerable low-degree nodes are themselves able to initiate the largest cascades.

Our theory, which is based only on degree correlations of connected nodes, accurately predicts the seed degree at which a local maximum in cascade size can occur in a real-world network: near the inverse threshold $\phi^{-1}$.  However, for the theory to correctly predict the cascade size itself, these networks must be rewired so as to randomize any structure beyond the joint degree distribution.
We know that local assortativity is more heterogeneous in real-world networks than in synthetic ones, with certain parts of a given network being assortative and other parts being disassortative. 
This suggests that manipulating a network's higher-order structure may allow us to tune the size of cascades, with an appropriate rewiring strategy offering a valuable tool for tailoring its stability to outbreaks.

\appendix
\section{List of Networks}
The six networks we study are from a variety of domains, including social networks (Facebook~\cite{snapnets}, Digg~\cite{snapnets}), biological (Reactome~\cite{snapnets}), Co-authorship (HepPh~\cite{snapnets}, HepTh~\cite{snapnets}), and Semantic networks (WordNet~\cite{wordnet}). The basic properties of networks we used in this paper are listed in Table~\ref{network}.
\begin{table}[H]
\begin{center}
\begin{tabular}{|l|l|r|r|r|r|}\hline
Network & Type & Nodes & Edges & $\kavg$ & Assort.\\\hline
Facebook & Social & 4,039 & 88,234 & 43.69 & 0.1660\\
Digg & Social & 27,567 & 175,892 & 12.76 & 0.1660\\
Reactome & Biological & 6,327 & 146,160 & 46.64 & 0.2449\\
ArXiv HepPh & Co-authorship & 12,008 & 118,489 & 19.74 & 0.6322\\
ArXiv HepTh & Co-authorship & 9,877 & 25,998 & 5.26 & 0.2679\\
WordNet & Semantic & 146,005 & 656,999 & 9.00 & $-0.0623$\\\hline
\end{tabular}
\end{center}
\caption{List of real-world networks and their basic profiles.}
\label{network}
\end{table}


\bibliography{references}

\begin{thebibliography}{17}%
\makeatletter
\providecommand \@ifxundefined [1]{%
 \@ifx{#1\undefined}
}%
\providecommand \@ifnum [1]{%
 \ifnum #1\expandafter \@firstoftwo
 \else \expandafter \@secondoftwo
 \fi
}%
\providecommand \@ifx [1]{%
 \ifx #1\expandafter \@firstoftwo
 \else \expandafter \@secondoftwo
 \fi
}%
\providecommand \natexlab [1]{#1}%
\providecommand \enquote  [1]{``#1''}%
\providecommand \bibnamefont  [1]{#1}%
\providecommand \bibfnamefont [1]{#1}%
\providecommand \citenamefont [1]{#1}%
\providecommand \href@noop [0]{\@secondoftwo}%
\providecommand \href [0]{\begingroup \@sanitize@url \@href}%
\providecommand \@href[1]{\@@startlink{#1}\@@href}%
\providecommand \@@href[1]{\endgroup#1\@@endlink}%
\providecommand \@sanitize@url [0]{\catcode `\\12\catcode `\$12\catcode
  `\&12\catcode `\#12\catcode `\^12\catcode `\_12\catcode `\%12\relax}%
\providecommand \@@startlink[1]{}%
\providecommand \@@endlink[0]{}%
\providecommand \url  [0]{\begingroup\@sanitize@url \@url }%
\providecommand \@url [1]{\endgroup\@href {#1}{\urlprefix }}%
\providecommand \urlprefix  [0]{URL }%
\providecommand \Eprint [0]{\href }%
\providecommand \doibase [0]{http://dx.doi.org/}%
\providecommand \selectlanguage [0]{\@gobble}%
\providecommand \bibinfo  [0]{\@secondoftwo}%
\providecommand \bibfield  [0]{\@secondoftwo}%
\providecommand \translation [1]{[#1]}%
\providecommand \BibitemOpen [0]{}%
\providecommand \bibitemStop [0]{}%
\providecommand \bibitemNoStop [0]{.\EOS\space}%
\providecommand \EOS [0]{\spacefactor3000\relax}%
\providecommand \BibitemShut  [1]{\csname bibitem#1\endcsname}%
\let\auto@bib@innerbib\@empty
\bibitem [{\citenamefont {Goltsev}\ \emph {et~al.}(2008)\citenamefont
  {Goltsev}, \citenamefont {Dorogovtsev},\ and\ \citenamefont
  {Mendes.}}]{Goltsev2008}%
  \BibitemOpen
  \bibfield  {author} {\bibinfo {author} {\bibfnamefont {A.~V.}\ \bibnamefont
  {Goltsev}}, \bibinfo {author} {\bibfnamefont {S.~N.}\ \bibnamefont
  {Dorogovtsev}}, \ and\ \bibinfo {author} {\bibfnamefont {J.~F.~F.}\
  \bibnamefont {Mendes.}},\ }\href@noop {} {\bibfield  {journal} {\bibinfo
  {journal} {Physical Review E}\ }\textbf {\bibinfo {volume} {78}},\ \bibinfo
  {pages} {051105} (\bibinfo {year} {2008})}\BibitemShut {NoStop}%
\bibitem [{\citenamefont {Karrer}\ \emph {et~al.}(2014)\citenamefont {Karrer},
  \citenamefont {Newman},\ and\ \citenamefont {Zdeborov\'a}}]{Karrer2014}%
  \BibitemOpen
  \bibfield  {author} {\bibinfo {author} {\bibfnamefont {B.}~\bibnamefont
  {Karrer}}, \bibinfo {author} {\bibfnamefont {M.~E.~J.}\ \bibnamefont
  {Newman}}, \ and\ \bibinfo {author} {\bibfnamefont {L.}~\bibnamefont
  {Zdeborov\'a}},\ }\href@noop {} {\bibfield  {journal} {\bibinfo  {journal}
  {Physical Review Letters}\ }\textbf {\bibinfo {volume} {113}},\ \bibinfo
  {pages} {208702} (\bibinfo {year} {2014})}\BibitemShut {NoStop}%
\bibitem [{\citenamefont {Granovetter}(1978)}]{Granovetter78}%
  \BibitemOpen
  \bibfield  {author} {\bibinfo {author} {\bibfnamefont {M.}~\bibnamefont
  {Granovetter}},\ }\href@noop {} {\bibfield  {journal} {\bibinfo  {journal}
  {American Journal of Sociology}\ }\textbf {\bibinfo {volume} {83}},\ \bibinfo
  {pages} {1420} (\bibinfo {year} {1978})}\BibitemShut {NoStop}%
\bibitem [{\citenamefont {Watts}(2002)}]{Watts2002}%
  \BibitemOpen
  \bibfield  {author} {\bibinfo {author} {\bibfnamefont {D.~J.}\ \bibnamefont
  {Watts}},\ }\href@noop {} {\bibfield  {journal} {\bibinfo  {journal}
  {Proceedings of the National Academy of Sciences}\ }\textbf {\bibinfo
  {volume} {99}},\ \bibinfo {pages} {5766} (\bibinfo {year}
  {2002})}\BibitemShut {NoStop}%
\bibitem [{\citenamefont {Centola}(2010)}]{Centola2010}%
  \BibitemOpen
  \bibfield  {author} {\bibinfo {author} {\bibfnamefont {D.}~\bibnamefont
  {Centola}},\ }\href {\doibase 10.1126/science.1185231} {\bibfield  {journal}
  {\bibinfo  {journal} {Science}\ }\textbf {\bibinfo {volume} {329}},\ \bibinfo
  {pages} {1194} (\bibinfo {year} {2010})}\BibitemShut {NoStop}%
\bibitem [{\citenamefont {Kempe}\ \emph {et~al.}(2003)\citenamefont {Kempe},
  \citenamefont {Kleinberg},\ and\ \citenamefont {Tardos}}]{Kempe2003}%
  \BibitemOpen
  \bibfield  {author} {\bibinfo {author} {\bibfnamefont {D.}~\bibnamefont
  {Kempe}}, \bibinfo {author} {\bibfnamefont {J.}~\bibnamefont {Kleinberg}}, \
  and\ \bibinfo {author} {\bibfnamefont {E.}~\bibnamefont {Tardos}},\
  }\href@noop {} {\bibfield  {journal} {\bibinfo  {journal} {KDD '03:
  Proceedings of the ninth ACM SIGKDD international conference on Knowledge
  discovery and data mining}\ ,\ \bibinfo {pages} {137}} (\bibinfo {year}
  {2003})}\BibitemShut {NoStop}%
\bibitem [{\citenamefont {Watts}\ and\ \citenamefont
  {Strogatz}(1998)}]{Watts1998}%
  \BibitemOpen
  \bibfield  {author} {\bibinfo {author} {\bibfnamefont {D.~J.}\ \bibnamefont
  {Watts}}\ and\ \bibinfo {author} {\bibfnamefont {S.~H.}\ \bibnamefont
  {Strogatz}},\ }\href@noop {} {\bibfield  {journal} {\bibinfo  {journal}
  {Nature}\ }\textbf {\bibinfo {volume} {391}},\ \bibinfo {pages} {440}
  (\bibinfo {year} {1998})}\BibitemShut {NoStop}%
\bibitem [{\citenamefont {Haldane}\ and\ \citenamefont
  {May}(2011)}]{haldane2011systemic}%
  \BibitemOpen
  \bibfield  {author} {\bibinfo {author} {\bibfnamefont {A.~G.}\ \bibnamefont
  {Haldane}}\ and\ \bibinfo {author} {\bibfnamefont {R.~M.}\ \bibnamefont
  {May}},\ }\href@noop {} {\bibfield  {journal} {\bibinfo  {journal} {Nature}\
  }\textbf {\bibinfo {volume} {469}},\ \bibinfo {pages} {351} (\bibinfo {year}
  {2011})}\BibitemShut {NoStop}%
\bibitem [{\citenamefont {Gleeson}(2008)}]{Gleeson2008}%
  \BibitemOpen
  \bibfield  {author} {\bibinfo {author} {\bibfnamefont {J.~P.}\ \bibnamefont
  {Gleeson}},\ }\href@noop {} {\bibfield  {journal} {\bibinfo  {journal} {Phys.
  Rev. E}\ }\textbf {\bibinfo {volume} {77}},\ \bibinfo {pages} {046117}
  (\bibinfo {year} {2008})}\BibitemShut {NoStop}%
\bibitem [{\citenamefont {Payne}\ \emph {et~al.}(2009)\citenamefont {Payne},
  \citenamefont {Dodds},\ and\ \citenamefont {Eppstein}}]{Payne2009}%
  \BibitemOpen
  \bibfield  {author} {\bibinfo {author} {\bibfnamefont {J.~L.}\ \bibnamefont
  {Payne}}, \bibinfo {author} {\bibfnamefont {P.~S.}\ \bibnamefont {Dodds}}, \
  and\ \bibinfo {author} {\bibfnamefont {M.~J.}\ \bibnamefont {Eppstein}},\
  }\href@noop {} {\bibfield  {journal} {\bibinfo  {journal} {Physical Review
  E}\ }\textbf {\bibinfo {volume} {80}},\ \bibinfo {pages} {026125} (\bibinfo
  {year} {2009})}\BibitemShut {NoStop}%
\bibitem [{\citenamefont {Newman}(2002)}]{Newman2002}%
  \BibitemOpen
  \bibfield  {author} {\bibinfo {author} {\bibfnamefont {M.~E.~J.}\
  \bibnamefont {Newman}},\ }\href {\doibase 10.1103/PhysRevLett.89.208701}
  {\bibfield  {journal} {\bibinfo  {journal} {Phys. Rev. Lett.}\ }\textbf
  {\bibinfo {volume} {89}},\ \bibinfo {pages} {208701} (\bibinfo {year}
  {2002})}\BibitemShut {NoStop}%
\bibitem [{\citenamefont {Dodds}\ and\ \citenamefont
  {Payne}(2009)}]{Dodds2009}%
  \BibitemOpen
  \bibfield  {author} {\bibinfo {author} {\bibfnamefont {P.~S.}\ \bibnamefont
  {Dodds}}\ and\ \bibinfo {author} {\bibfnamefont {J.~L.}\ \bibnamefont
  {Payne}},\ }\href@noop {} {\bibfield  {journal} {\bibinfo  {journal}
  {Physical Review E}\ }\textbf {\bibinfo {volume} {79}},\ \bibinfo {pages}
  {066115} (\bibinfo {year} {2009})}\BibitemShut {NoStop}%
\bibitem [{\citenamefont {Wu}\ \emph {et~al.}(2017)\citenamefont {Wu},
  \citenamefont {Percus},\ and\ \citenamefont {Lerman}}]{Wu2017}%
  \BibitemOpen
  \bibfield  {author} {\bibinfo {author} {\bibfnamefont {X.-Z.}\ \bibnamefont
  {Wu}}, \bibinfo {author} {\bibfnamefont {A.~G.}\ \bibnamefont {Percus}}, \
  and\ \bibinfo {author} {\bibfnamefont {K.}~\bibnamefont {Lerman}},\
  }\href@noop {} {\bibfield  {journal} {\bibinfo  {journal} {Scientific
  Reports}\ }\textbf {\bibinfo {volume} {7}},\ \bibinfo {pages} {5576}
  (\bibinfo {year} {2017})}\BibitemShut {NoStop}%
\bibitem [{\citenamefont {Boguna}\ \emph {et~al.}(2004)\citenamefont {Boguna},
  \citenamefont {Pastor-Satorras},\ and\ \citenamefont
  {Vespignani}}]{Boguna2004}%
  \BibitemOpen
  \bibfield  {author} {\bibinfo {author} {\bibfnamefont {M.}~\bibnamefont
  {Boguna}}, \bibinfo {author} {\bibfnamefont {R.}~\bibnamefont
  {Pastor-Satorras}}, \ and\ \bibinfo {author} {\bibfnamefont {A.}~\bibnamefont
  {Vespignani}},\ }\href {\doibase 10.1140/epjb/e2004-00038-8} {\bibfield
  {journal} {\bibinfo  {journal} {Eur. Phys. J. B}\ }\textbf {\bibinfo {volume}
  {38}},\ \bibinfo {pages} {205} (\bibinfo {year} {2004})}\BibitemShut
  {NoStop}%
\bibitem [{\citenamefont {Mahadevan}\ \emph {et~al.}(2006)\citenamefont
  {Mahadevan}, \citenamefont {Krioukov}, \citenamefont {Fall},\ and\
  \citenamefont {Vahdat}}]{Mahadevan2006}%
  \BibitemOpen
  \bibfield  {author} {\bibinfo {author} {\bibfnamefont {P.}~\bibnamefont
  {Mahadevan}}, \bibinfo {author} {\bibfnamefont {D.}~\bibnamefont {Krioukov}},
  \bibinfo {author} {\bibfnamefont {K.}~\bibnamefont {Fall}}, \ and\ \bibinfo
  {author} {\bibfnamefont {A.}~\bibnamefont {Vahdat}},\ }\href {\doibase
  10.1145/1151659.1159930} {\bibfield  {journal} {\bibinfo  {journal} {SIGCOMM
  '06 Proceedings of the 2006 conference on Applications, technologies,
  architectures, and protocols for computer communications}\ ,\ \bibinfo
  {pages} {135}} (\bibinfo {year} {2006})}\BibitemShut {NoStop}%
\bibitem [{\citenamefont {Leskovec}\ and\ \citenamefont
  {Krevl}(2014)}]{snapnets}%
  \BibitemOpen
  \bibfield  {author} {\bibinfo {author} {\bibfnamefont {J.}~\bibnamefont
  {Leskovec}}\ and\ \bibinfo {author} {\bibfnamefont {A.}~\bibnamefont
  {Krevl}},\ }\href@noop {} {\enquote {\bibinfo {title} {{SNAP Datasets}:
  {Stanford} large network dataset collection},}\ }\bibinfo {howpublished}
  {\url{http://snap.stanford.edu/data}} (\bibinfo {year} {2014})\BibitemShut
  {NoStop}%
\bibitem [{\citenamefont {Fellbaum}\ and\ \citenamefont
  {Tengi}(2005)}]{wordnet}%
  \BibitemOpen
  \bibfield  {author} {\bibinfo {author} {\bibfnamefont {C.}~\bibnamefont
  {Fellbaum}}\ and\ \bibinfo {author} {\bibfnamefont {R.}~\bibnamefont
  {Tengi}},\ }\href@noop {} {\enquote {\bibinfo {title} {Wordnet: A lexical
  database of english},}\ }\bibinfo {howpublished}
  {\url{http://wordnet.princeton.edu/}} (\bibinfo {year} {2005})\BibitemShut
  {NoStop}%
\end{thebibliography}%
\end{document}